\documentclass{lqz_natureprintstylev4}
\usepackage{astjnlabbrev-nature}
\usepackage{hyperref}

\usepackage{amssymb}
\usepackage{amsmath}	
\usepackage{mathtools}
\usepackage{gensymb}
\usepackage[svgnames]{xcolor}
\usepackage{enumitem}
\usepackage{textcomp}

\usepackage{epsfig}
\usepackage{graphicx}
\usepackage{longtable}
\usepackage{hyperref}

\usepackage[switch]{lineno}
\usepackage[labelsep=endash]{caption}

\def\etal{\textit{et~al.}}

\newcommand{\msun}{\ensuremath{M_{\odot}}}
\newcommand{\mbh}{\ensuremath{M_\mathrm{BH}}}

\newcommand{\elegantrule}{%
	\par\vspace{2ex}  % blank upper
	\noindent%
   {\centering
 	 \raisebox{0.8ex}{\textcolor{gray!40}{\rule{0.3\linewidth}{0.2pt}}}% left, thin line, raised by 0.8ex
 	 \hspace{0.02\linewidth}%
 	 \textcolor{gray!80}{$\hbar$}% center, ?
	 \hspace{0.02\linewidth}%
	 \raisebox{0.8ex}{\textcolor{gray!40}{\rule{0.3\linewidth}{0.2pt}}}% right, thin line, raised by 0.8ex
	 \par\vspace{1ex}% blank bottom
   } % end of centering
} %% elegant divider line.

\newcommand{\rev}[1]{{#1}}  %%2017Feb3 to cancel the highlight

\newcommand{\revMar}[1]{{#1}}

\newcommand{\revApr}[1]{{#1}}

\newcommand{\revMay}[1]{{#1}}

\newcommand{\ReasoningChainMar}[1]{{}}

\newcommand{\citep}{\cite}
\newcommand{\citet}{\cite}

\title{A Promise for the JWST era: Massive black holes directly collapsed from wave dark matter haloes, 
	and Star formation in and around their accretion flows}

\author{Xiaobo~Dong$^{1\,\dag}$, Yongda~Zhu$^{2\,\dag}$, Marcia~J.~Rieke$^{2}$,
	George~H.~Rieke$^{2}$, Xinyu~Li$^{3}$,
	Peter~Behroozi$^{2}$, Haixia~Ma$^{4}$, Runyu~Meng$^{1}$, 
        Zhiying~Mao$^{5}$, and Zhe~Sun$^{6}$ }

\begin{document}

\maketitle

\begin{affiliations}
\item Yunnan Observatories, Chinese Academy of Sciences, Kunming 650011, China;\\
\item Steward Observatory, University of Arizona, 933 North Cherry Avenue, Tucson, AZ 85721, USA;\\
\item Department of Astronomy, Tsinghua University, Beijing 100084, China;\\
\item Division of Particle and Astrophysical Science, Nagoya University, Nagoya, Aichi 464-8601, Japan;\\
%% \item University of Chinese Academy of Sciences, Shijingshan District, Beijing 100049, China; \\
\item INAF-Osservatorio di Astrofisica e Scienza dello Spazio di Bologna, Via Piero Gobetti 93/3, 40129 Bologna, Italy; \\
\item CAS Key Laboratory of Strongly-coupled Quantum Matter Physics, National Synchrotron Radiation Laboratory, University of Science and Technology of China, Hefei 230026, China;\\
$^{\dag}$ These authors contributed equally to this work.\\
\end{affiliations}

%\noindent { \textit{ver. 0.5(2024Dec24)} } \\
%\noindent { \textit{ver. 0.618(2025Feb28)} } \\
%\noindent { \textit{ver. 0.954(2025Apr2)} } \\
%\noindent { \textit{ver. 0.997(2025Apr23)} } \\

%\noindent
%{\textbf{\large One-Sentence Summary:}	
%	Wave CDM is not falsified by observations of galaxies, but can explain all puzzles so far.
%}
%\\

\begin{abstract}

%2024Oct31,cleaned-draft version.\\
%2024Nov2night, draft to YD, RY, HX, for a first read. \\
%2024Dec17noon (1st decent ver), to YD, RY, HX, ms for the 2nd zoom-meeting. \\

%Ver: \today
%\\

%%%%%%%%%%%%%%%%% Purpose:
%% //Einstein 1905:  to explain several puzzling phenomena
%      by assuming that blablabla ......
%   ---> integrate them into a coherent physical theory!
%%%%%%%%%%%
%$<$the interaction of light with matter consists of the emission or absorption of such light quanta.$>$ \\
% - Einstein constantly reminded his colleagues of the need to introduce 
%\textbf{ radically new concepts } 
%to explain the structure of both matter and radiation. 
%He himself introduced some of these new concepts, notably the light quantum hypothesis, 
%although he remained unable to integrate them into a coherent physical theory.
%%%%%%%%%%%%%%%%%%%%%%%%%%%%%%%%%%%%%%

There are several puzzling phenomena %(or even called tensions) 
from recent JWST observations,
which seem to push the standard $\Lambda$CDM cosmology 
(CDM standing for cold dark matter and $\Lambda$ being Einstein's cosmological constant) over the edge. 
All those puzzles can be understood in a coherent way 
if we assume that first massive black holes (MBHs, the ``heavy eggs'') formed 
by directly collapsing from  
wave CDM haloes,\cite{Chavanis2016,Helfer2017,Sikivie-Zhao2024} 
\revMar{which can be even} 
earlier than the formation of first galaxies (the ``chickens'').
We elucidate two false obstacles that have been believed to prevent wave CDM haloes from collapsing into black holes 
(namely ``maximum stable mass'' and ``bosenova'')
and point out that general-relativistic instabilities 
(e.g., the non-axisymmetric instability numerically found in spinning scalar-field boson stars\cite{Sanchis-Gual2019}) 
could serve as the concrete mechanisms for direct-collapse black holes (DCBHs) born from wave dark matter.
%%% After Hai.FU's comments, 2025Apr27: --
Once the MBHs formed, star formation bursts 
in\cite{Collin-Zahn1999} and around\cite{Silk-Rees1998,Collin-Zahn1999,Silk2024} the accretion flows,
characteristic of a special mode in compact regions of high gravitational accelerations.\cite{Silk2024,Boylan2024}

\end{abstract}

\begin{figure*}
	\centerline{\includegraphics[scale=0.22,angle=0]{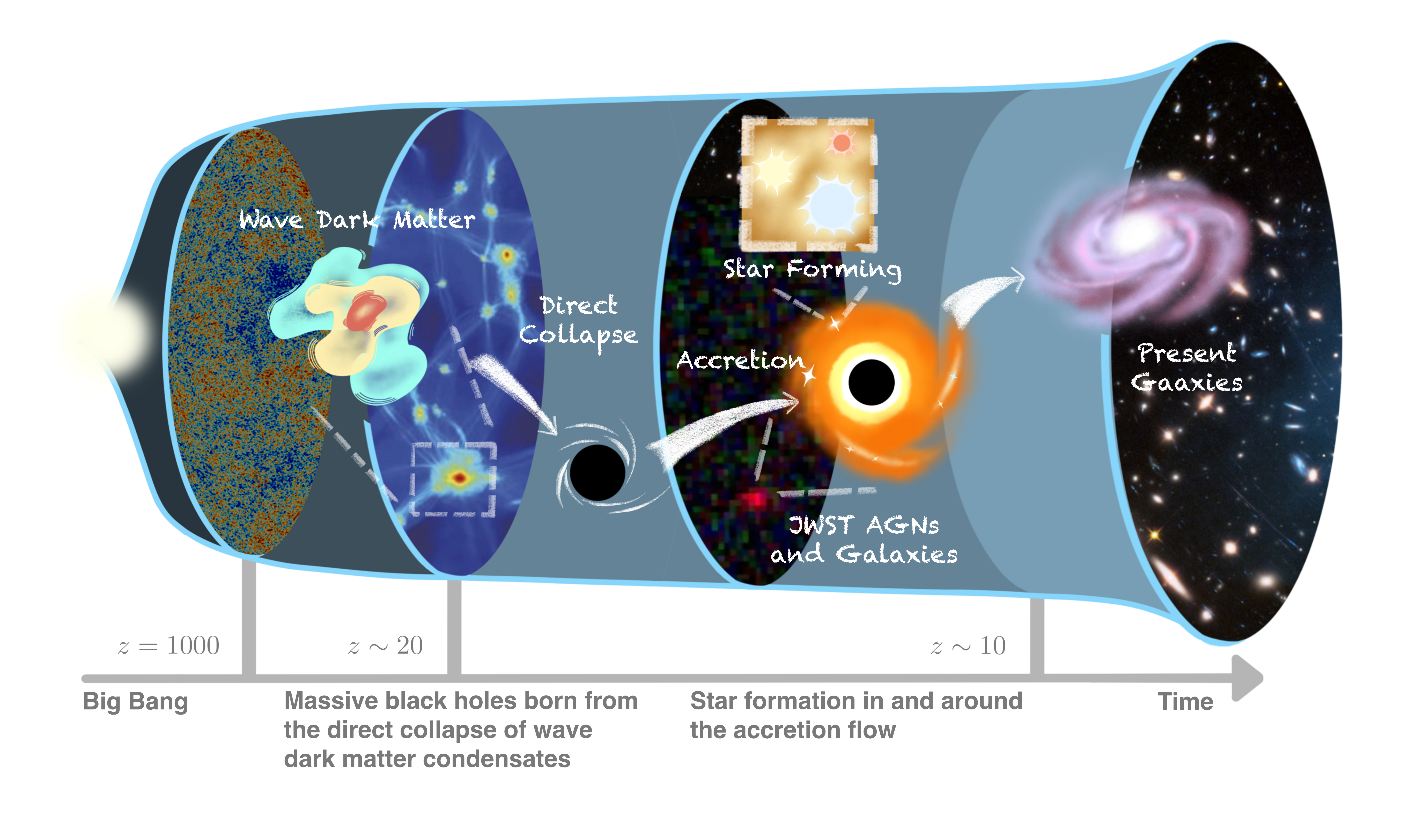}}
		%%{20250401_Haixia_2D3D-Figure1.pdf}}
	%\centerline{\includegraphics[]{Schematic_3D_Runyu_2024Dec22.eps}}
	\caption{\textbf{Schematic diagram of the theory proposed to explain the JWST puzzles and other ``cosmic gravitational conundrums.''} 
	The following three ingredients are integrated into a unified scenario:
	(1) the formation of massive black holes precedes first-generation galaxies\cite{Silk-Rees1998},
	(2) which are directly collapsed from wave dark matter 
	due to non-axisymmetric instability numerically found in spinning scalar-field boson stars\cite{Sanchis-Gual2019},
	and (3) the inside-to-outside star formation in and around the accretion flows toward massive black holes\cite{Silk-Rees1998,Collin-Zahn1999,Ali-Dib-Doug2023,JMWang2023,Silk2024}.					
		%%(draft ver.1; 2D+3D). 
		Data credits: CMB -- Bennett \etal\ 2013, Astrophys. J. Suppl. 208, 20; 
		waveDM LSS: -- Schive \etal\ 2014, Nat. Phys. 10, 496; 
		JWST image: Robertson \etal\ 2023, Nat. Astron. 7, 611; 
		%% waveDM granules -- Mocz \etal\ 2019, Phys. Rev. Lett. 123, 141301;
		HST image: NASA/ESA/Hubble, F.~Pacaud, D.~Coe\,.
	 }
	\label{final-Figure1}
\end{figure*}

%% \section*{Main text:}
%% Once the MBHs formed, the ensuing star formation processes in and around the accretion flows, 
%% 2025Apr27 --:
The above starburst processes, as well as the accretion activity \textit{per se} dubbed active galactic nulei (AGNs),
can naturally explain all the following phenomena:
the occurrence of JWST's \emph{dense and compact} ``little red dots'' (galaxies or AGNs), 
the mature appearance of \emph{dense and compact} galactic components such as massive disks and bulges at high redshifts,
the over-massive supermassive black holes (SMBHs), 
and the pattern of the spatial distribution of metallicity
in high-redshift galaxies and quasars,  
namely \emph{higher in the AGN broad-line regions (BLRs) 
	than in the narrow-line regions} (NLRs, broadly the ISM of the host galaxies).
The sequence of the proposed processes and some related phenomena 
are illustrated in Figure~\ref{final-Figure1}.
%%% 
Also, this proposal can explain the enhanced star formation efficiency 
and even the enhanced small-scale baryonic fraction
required by the abundant massive galaxies at cosmic dawn (the re-ionization epoch).

%% Old text, for check: 
%% Once the MBHs formed, the ensuing processes in and around the accretion flows toward the black hole (i.e., the AGN) 
%% can naturally explain all the JWST puzzles including the enhanced star formation efficiency and xxxx
%% concerning compact and massive galaxies (JWST's little red dots),
%% over-massive black holes, 
%% and the spatial pattern of metallicity in galaxies and quasars at cosmic dawn.
%% These processes include 
%%  the star formation activity triggered by the positive feedback of AGN outflows 
%%  \cite{Silk-Rees1998,Collin-Zahn1999,Silk2024},
%%  %(SilkRees98; CollinZahn99; Silk+2024),
%%  and the route that stars can be born and die in the accretion flows
%%   \cite{Collin-Zahn1999,Ali-Dib-Doug2023,Huang-Doug-Shields2023},
%%   %(CollinZahn99; AliDib-Doug23,Huang-Doug-Shields23).
%% ---------------.

%The merit of this paper: to integrate the above points of multiple disciplines 
%into a coherent physical theory.
%%%
%Several SPARKs.
% https://ui.adsabs.harvard.edu/abs/1995ApJ...443...11E/abstract
% https://ui.adsabs.harvard.edu/abs/1998A%26A...331L...1S/abstract
% https://ui.adsabs.harvard.edu/abs/1999A%26A...344..433C/abstract
%\textbf{Old ideas yet radical still.}
%

Interestingly, looking in this way, we will see below that 
several independent ideas from three decades ago%{EisensteinLoeb95,SilkRees98,CollinZahn99}
\cite{Eisenstein-Loeb1995,Silk-Rees1998,Collin-Zahn1999}
(before the construction of JWST and parallel with its early development of the ``NGST at the 6--8-m level'' concept)
%% disparate ideas from three decades ago 
--- \emph{sparks old but ``radical'' still} --- 
can now be linked with the recent discoveries of JWST
and may together be integrated into a coherent physical theory (see below). % \\

\section*{JWST puzzles}
%%  --- OUTLINE and incremental words: --- 
%%pick up several most related puzzling phenomena from the recent review \cite{Adamo-JWST2024} 
%%and some other recent papers:
%%%
%1. the biggest one -- the ``impossibly early formation problem'',
%more abundant luminous galaxies at $z > 7$ than conventional $\Lambda$CDM prediction, 
%enhanced star formation rate \cite{Boylan2023}.

The first two years of JWST observations have unlocked a Pandora's box of 
puzzling discoveries about galaxies and SMBHs in the early Universe, 
challenging our previous understanding based on the standard $\Lambda$CDM cosmological model
(see \cite{Adamo-JWST2024} for a recent review).
%%%
From the perspective of astronomical observables,
these recent puzzles, together with the long-standing mysteries of
dark energy, dark matter and the origin of SMBHs, 
%% all involve gravity 
are in fact all fundamentally linked to gravity 
and thus can be collectively referred to as ``cosmic gravitational conundrums.''
Below, we outline the major JWST puzzles, for which we will provide a unified physical explanation. \\

\noindent
\textbf{1. ``Impossibly early formation problem.''}\\
The biggest puzzle is undoubtedly the ``impossibly early formation problem,'' 
as highlighted in many research papers entitled like this or similar 
(accelerated early galaxy formation, enhanced matter power spectrum on galactic scales, etc.),
discovering more abundant luminous galaxies at $z > 7$ than conventional $\Lambda$CDM prediction.\cite{Adamo-JWST2024,Boylan2023} 
Broadly speaking, this early formation of massive galaxies is quite unusual
in the $\Lambda$CDM paradigm that predicts a ``bottom-up'' formation scenario for galaxies,
where it is dwarf galaxies, rather than massive ones, that could be so numerous during the re-ionization epoch.
Second, this requires an unusually high star formation efficiency converting baryonic gas into stars.
Finally and probably most severely,
the cosmic stellar mass density of those galaxies \rev{seems} %is - 2025Feb22 
so high as to exceed the maximum amount of all baryons in the related dark matter haloes %%.\cite{Boylan2023,Boylan2024} 
\revApr{(see, e.g., refs.\cite{Boylan2023,Boylan2024}; but cf. \cite{Woodrum2024} for the uncertainties 
caused by IMF by a factor of three or even more).
%% below, revised 2025Apr15:
%% Nevertheless, we note that 
Some recent, less model-dependent observations seem still to suggest potential ``early formation.''
The latest determination of the ultraviolet luminosity function of JWST galaxies reveals 
a systematic discrepancy higher than theoretical expectations at $z \gtrsim 12$ at all luminosities.\cite{Whitler2025}
%% Yongda's words about latest ALMA and extrme overdensity (his comment on Overleaf:)
Ref.\citet{2502.06016} reports a kinematic gas mass of $\log(M_{\rm gas}/M_\odot) = 9.8 \pm 0.3$ 
in a galaxy at $z = 14$ based on ALMA observations.  
Ref.\citet{2503.15597} identifies 
an extreme galaxy overdensity at $z \sim 8.47$ using NIRCam/WFSS data; 
comparisons with cosmological simulations indicate that 
the probability of finding such a large-scale structure is less than 5\% 
within the current galaxy formation framework and the observed survey volume.
Thus this tension is likely to be nailed down in the near future.  
} %\revApr ; 2025Apr

%% new para:
Also in such an ``impossibly early'' situation is the number density of active MBHs 
(namely AGNs, including quasars) discovered at $z > 7$ (e.g., \cite{PacucciApJL2023}).

%% Pacucci_B.Nguyen_S.Carniani_Roberto.Maiolino_Xiaohui.Fan_2023ApJL__JWST CEERS and JADES 
%%   Active Galaxies at z=4--7 Violate the Local M_bh -- M_galaxy Relation 
%%  at gt 3 --- Implications for Low-mass Black Holes and Seeding Models

The key point is that the ``impossibly early formation'' is all-around and systematic, 
manifesting not only in the abundance of quasars and massive galaxies at cosmic dawn 
but also in the early formation of mature galactic components 
including massive disks\cite{Nelson-disk2023} and 
(proto-)bulges\cite{Baker-bulge2024}. %(Baker+2024NatAstron).
A striking characteristic in common among those galaxies and galactic sub-structures is 
that they are \emph{both compact and dense}. 
\newline

% ---MEMO: to give the answer in the inside-to-outside SF section; \\
% the enhanced gravity by waveDM and first quasars; 
% on galactic and smaller scales, cf. Boylan2024's proposal (in CDM). \\

%% W.M.Baker_Sandro.Tacchella_etal_2024Oct_NatAstron_
%% A core in a star-forming disc as evidence of inside-out growth in the early Universe (JWST JADES; proto-bulge)

%% 2310.06887v1_E.Nelson_G.Brammer+_ApJ__FRESCO - An extended, massive, rapidly rotating galaxy at z=5.3 JWST

\noindent
\textbf{2. Over-massive black holes, and the possible presence of naked quasars
before the birth of first stars and first galaxies.}\\
%% Evidence for heavy-seed origin of early supermassive black holes from a z~10 X-ray quasar
%% Akos Bogdan 1 , Andy D. Goulding2,10 , Priyamvada Natarajan  et al.
%%
JWST observations have revealed a number of high-redshift AGNs with black hole masses 
that appear 10--100 times larger than expected from the stellar mass ($M_\star$) of their host galaxies
in terms of the local $M_\mathrm{BH}$--$M_\star$ scaling relation.\cite{Adamo-JWST2024} 
Some of those SMBHs are comparable to or probably even more massive than 
their host galaxies.\cite{Bogdan-UHZ1-2024,Chen2024} 
%%%
%% Although recently, \citet{Sun2024} suggest that this relation does not change significantly up to $z \sim 4$, 
%% data from higher redshifts suggests otherwise. 
For instance, quasars at $z \sim 6$ observed with JWST/NIRCam %%\citep{Rieke2023} 
exhibit \mbh\ values around 10 per cent of $M_\star$, 
% $M_\mathrm{BH}/M_\star$ ratios around $\log(M_\mathrm{BH}/M_\star) \sim -1$, 
notably higher than their lower-redshift counterparts 
in which \mbh\ are approximately 0.3 per cent of $M_\star$.\citep{Ding2023} 
%$\log(M_\mathrm{BH}/M_\star) \sim -2.5$  %\citep{Ding2023, Yue2024, Stone2024}. 
%
This redshift evolution in the relation has also been noted for smaller SMBHs in Seyfert-luminosity AGNs 
at $4 < z < 7$\,\cite{PacucciApJL2023,Harikane2023} 
%% \citep{Ubler2023, Maiolino2023, Harikane2023, PacucciApJL2023}. 
%% -- Chen,Ho etal 2024 --:
Furthermore, ref.\cite{Chen2024} analyzed the images of eight little red dots (LRDs, belonging to AGNs or galaxies) 
detected by JWST, and found that none but one has reliable evidence for the presence of a host galaxy;
either the measured stellar mass or the upper limits 
are on the same order of magnitude as or even lower than the masses of the SMBHs.
%% with stellar masses much lower than the expectation 
%%based on the black hole masses from broad H$\alpha$ emission line and the local $M_\mathrm{BH}$--$M_\star$ scaling relation. 
%% These observations imply an emerging population of black holes 
%% that are more massive than their host galaxies in the early universe, posing a challenge to black hole-galaxy coevolution models.
%%
This suggests the possible presence of naked SMBHs and naked quasars in the early Universe, before the birth of first stars and first galaxies.
%% If confirmed by future observations of deeper imaging and spectroscopy ....
\newline

\noindent
\textbf{3. Unexpected rapid chemical enrichment in galaxies and quasars at $z>7$.} \\
%%
%%( for galaxies, as early as $z=12.5$, only 350 Myr after the Big Bang)
%%//Here, three questions for YD: Q1. what is the situation of the two z=14 galaxies? \\
%%%
%%YD 2024.11.8: update of the metallicity of the two z=14 galaxies. \\
%%Key point: no first stars, no first galaxies!
% updated version below -YD 2024Nov10; xbdong 2024Dec4night:
%
The studies prior to JWST indicated that the rest-UV continuum slopes of early star-forming galaxies are typically blue, 
consistent with young, dust-poor systems. 
However, after launch JWST has soon revealed a population of dusty, optically faint sources that were previously undetected.\cite{Adamo-JWST2024} 
For example, results from the CEERS survey\cite{Perez-Gonzalez2023} have identified 
a diverse array of dusty star-forming galaxies, quiescent galaxies, and compact starbursts at $z > 6$, 
with significant dust attenuation and high star formation rates, 
suggesting that dust and metal enrichment processes were already well-established by this epoch. 
Spectroscopic confirmation of the two earliest galaxies at $z \approx 14$.~\cite{Carniani2024, Robertson2024} 
shows evidence of advanced metal enrichment with $Z/Z_\odot > 0.01$, 
along with detectable dust, reinforcing the idea that 
galaxies in the early Universe underwent rapid chemical evolution within only 300 million years after the Big Bang. 
For the one called \mbox{GS-z14-0}, ref.\cite{Schouws2024} detected the \mbox{[O\,III]} 88$\mu$m emission line, 
indicating a metallicity of $Z \approx 0.05-0.2 \, Z_\odot$ even higher than the estimate at first.
%% which supports the notion that early galaxies underwent rapid metal enrichment.
To conclude, the non-detection by JWST of first stars and first galaxies, 
which should be metal-free and expected as Population III (Pop III) stars, 
underscores the surprising efficiency and early onset of metal enrichment processes 
in the first few hundred million years after the Big Bang.

On the side of AGNs (including quasars),
it has been well known for more than 30 years\cite{Hamann-Ferland1999} that 
high-redshift quasars (at $z < 5$ before the year of 1999) typically exhibit quite high metallicities,
solar or even several times larger, in the central few parsecs around the SMBHs,
as probed by both intrinsic absorption lines and broad emission lines from the broad line regions (BLRs).
%%%
%% In addition to galaxies, AGNs at $z \sim 6-7$, such as the highly reddened Abell2744$-$QSO1, 
%% display significant dust content and rapid black hole growth \cite{Furtak2024} (see also \cite{Christensen2023} by JWST), 
%% indicating that metal and dust enrichment processes were active even in early SMBH environments. 
%%%%
%% Consistent with this trend, recent studies report no redshift evolution in the Fe II/Mg II flux ratios of quasars 
%% from $z = 1$ to $6.6$ \citep{Jiang2024}, supporting the idea that rapid enrichment processes operated early on.
%%%
Right now, JWST observations of quasars at higher redshifts, including sources like J1120$+$0641 at $z > 7$, 
reveal BLRs with metallicities comparable to those of later cosmic times,\cite{Bosman2024}  
as well as dusty tori, suggesting that the surprisingly early enrichment is tracked back to the cosmic dawn.

%% the most crucial point: -- 
A revelational point may lie in the spatial pattern of metallicity around quasars:
the metallicities in the AGN BLR are higher than in the narrow-line region 
(see ref.\cite{Huang-Doug-Shields2023} and the references collected therein).
In an AGN, the BLR is near the SMBH within 1 parsec, roughly on the scale of the self-gravitating part of the accretion flow.
The NLR is at least outside the dusty torus (beyond 1 parsec), and basically belongs to the inter-stellar medium of the host galaxy.
Hence this fact suggests that metals are created near the SMBHs.\cite{Silk-Rees1998,Collin-Zahn1999}
\newline

%%{Star formation: ``Inside-to-Outside'' processes in and around the accretion flows}
% (1) SF inside accretion flows: collinZahn99, Doug.Lin; also SilkRees98.
% (2) to outside SF:  by SNe shock: CollinZahn99; by AGN outlow shock: SilkRees98, Silk+2024.
%
%%%%
%% From the perspective of astronomical observables,
%% dark energy and dark matter, the seed of SMBHs, and the puzzles from recent JWST observations, 
%% all involve gravity 
%% and can be collectively termed ``cosmic gravitational conundrums.''

In summary, as stressed in the JWST-based papers cited above, 
(1) the impossibly early formation of massive galaxies and mature galactic sub-structures,
(2) the origin of SMBHs and possible presence of naked quasars at the cosmic dawn,
and (3) the presence of non-pristine gas (metal-enriched) at $z=14$,
along with the non-detection of Pop. III stars and first-generation galaxies so far, 
are each difficult to account for in the conventional $\Lambda$CDM paradigm,
let alone to explain all of them at one fell swoop.
Below we wish to present our train of thought aiming at a coherent theory
based on some recent developments of multiple disciplines, 
in the hope that the idea would be useful for researchers in their investigations.

%%2024Dec24noon, check pauses here, -xbdong.

\section*{wave CDM}
%% L.Hui 2021ARAA:
% a good dark matter candidate should be cold and weakly interacting.
% both weak self-interaction and weakly coupling with ordinary matter 
% %%%
%Wave dark matter,such as the axion, is in fact one form of cold dark matter, CDM (fuzzy CDM, Hu+2000-fuzzyCDM). 
%We use the term particle dark matter for cases in which m  30 eV, the primary example of which
%is a weakly interacting massive particle (WIMP). We sometimes refer to it as conventional CDM.
%
%% Hu+2000 fuzzy CDM:
% While interesting, these solutions require self-interactions wildly in excess of
% those expected for weakly interacting massive particles or axions.
%
According to the principle of Occam's razor,
we can minimally make just a single adjustment to the standard cosmological model:
from massive particle CDM to ultralight wave CDM %(Hu+2000-fuzzyCDM, Hui2010waveCDM),
\cite{Hu2000,Niemeyer2020,Ferreira2021,Hui2021}, 
which in fact was proposed decades ago to account for so-called ``small-scale crises'' 
concerning the (circum-)galactic scales ($\lesssim$ 100 kpc).
%%%%
By requiring both wave (ultralight) and cold, 
the dark matter must be bosonic.
%%% collisionless:
Furthermore, to be consistent with the astronomical observations, 
the dark matter should be deemed collision-less; 
to be exact, we assume that both the self interaction and 
the coupling with normal matter are weaker than gravitation 
(cf. \cite{JiaoBrandenbergerKamali2025} for an opposite proposal). 
Here we do not consider the hybrid models that dark matter is 
a mixture of both collisionless and collisional types\cite{Pollack2015},
nor the models with self-interaction in some velocity-dependent fashion (e.g., \cite{F.Jiang2025}). 
%% Jason.Pollack_David.Spergel_Paul.Steinhardt_2015ApJ,  2015ApJ...804..131P: 
%% Supermassive Black Holes from Ultra-Strongly Self-Interacting Dark Matter.
%%  : assuming a small fraction ( f <~ 0.1 ) of the dark matter is ultra-strongly self-interacting
%%  :the non-SIDM component is taken to be collisionless, has s = 0.
%%%%
In summary, we still work within the framework of CDM,
and do not require primordial black holes or other forms of dark stuff/energy 
that have existed at $z>1100$ 
before the epoch of decoupling of normal matter from photons
(except for the above ultralight bosonic field). %%\cite{Hui2021}).
For this reason, in this article we purposely use the term ``wave CDM''
(in the same sense as ref.\cite{Hu2000} used ``Fuzzy CDM''),  
which should be understood being interchangeable with ``wave DM.''
%%  i.e., before the epoch of recombination.
%%  Cosmic Microwave Background (CMB) is the relic of the decoupling.
%%%%%%
Thus the major merit is that the initial conditions of wave CDM,
for most astronomical purposes such as the large-scale structure and galaxy formation and evolution,
is well constrained by CMB observations, 
and can be set essentially in the same way as 
the routine practice working in the conventional particle CDM paradigm 
\rev{(i.e., no need for the information at $z>1100$)}. %2022Feb22
A second merit is that this model has unique observational signatures to test directly,
i.e., ready to be falsifiable soon.

%% 2024.11.30:
Note that at this point we do not prescribe the exact mass of the bosons.
This is because astrophysical probes are generally affected by rich physical processes 
(e.g., the wave kinetics and relaxation, and even the highly speculated genesis and freeze-out mechanisms)\cite{Niemeyer2020} 
rather than solely by the boson mass 
(e.g., the sources of systematic uncertainties listed in ref.\cite{Hui2021}; 
\mbox{R.Meng,\,X.Dong,\,\etal, in prep.}; 
\revMar{see also similar comments in \mbox{refs.\cite{MaySpringel2023,Widmark2024,ChanSchive2025}}}).
To think positive, instead,
this means that those astrophysical phenomena is not very sensitive to the mass of DM particles,
which is in fact the very essence of the CDM paradigm generically, 
\revMar{be particles or ultralight fields, be fermions or bosons,}
and the reason of its long phenomenological success.

%% new para: 2024Dec11:
Wave CDM bears a distinctive characteristic:
\emph{both} enhancing the overall gravity on galactic and sub-galactic scales 
due to Bose-Einstein condensation (BEC)\cite{Ferreira2021} and wave interference (in contrast with warm dark matter)\cite{Hui2021}
\emph{and} blurring the detailed structures of matter on those so-called small scales
due to quantum pressure (in contrast with both conventional particle CDM and modified-gravity theories)\cite{Hu2000}.
This two-fold characteristic is crucial,
from the standing point of astronomical observations.
 
%//---------- Note: \\
%Two basic properties: ultralight + bosonic. 
%In this sense, the discussions here apply to  dark photons (bosons with spin $s=1$)
%and any other ultralight bosonic DM models.
%But the analysis of GR instabilities would be more complicated \cite{Sanchis-Gual2019}.
%

\section*{Ready cold matter with zero angular momentum}

Discoveries of accreting SMBHs at high redshifts, 
\rev{
even about 400 million years after the Big Bang\cite{Maiolino2024_GN-z11},
have increasingly provided evidence to support 
that they originate from `heavy' seed BHs with masses above $\sim10^4$ solar mass.\cite{Bogdan-UHZ1-2024} 
} %2025Feb22 
Conventional heavy seed theories are based on 
the direct collapse of high-redshift baryonic gas in special conditions,
being rendered low angular momentum,
and normally being required 
both (1) free of metals and Hydrogen molecules 
and (2) free of star formation and feedback activities.\cite{Inayoshi2020,Zwick2023,Mayer2024,JiaoBrandenbergerKamali2025}

However, there is a common difficulty of the conventional direct collapse scenarios:
a large reservoir of pre-existing cold baryonic gas with almost zero angular momentum
as a prerequisite condition, no matter neither with galaxy--merger driving\cite{Zwick2023,Mayer2024} or without
(e.g., \cite{Eisenstein-Loeb1995, Loeb2024}; see the recent review\cite{Inayoshi2020} for other concrete scenarios).
%%%
These scenarios involving matter with almost zero angular momentum 
is in fact the concrete implementation of the so-called ``coherent collapse'' supposed by ref.\cite{Silk-Rees1998}\,.

% new para: 
In the wave CDM scenario, 
\rev{the temperature of bosons in the Universe 
	is many orders of magnitude below the critical temperature ($T_\mathrm{c}$) required for their BEC phase transition.\cite{Ferreira2021,Hui2021}} 
Thus the matter with zero angular momentum is already ready plentifully: 
the Bose-Einstein condensates with momentum $k \rightarrow 0$ ($k$ denoting wave vector)
such as the soliton-like cores, 
which are the inevitable components of wave CDM haloes 
naturally formed in wave kinetic processes as a type of coherent structures.\cite{Niemeyer2020,Hui2021} 
%%% 2025Feb20 added below:
%\rev{}
The bulk of a condensate -- excluding quantized vortex lines (a type of topological defect) within it 
	-- cannot sustain rotational motion;
	its bulk velocity field is curl-free anywhere and carries precisely zero bulk angular momentum
	with respect to any reference frame.
%% 2025:	
In realistic cosmic DM haloes, angular momentum is generally present. 
However, this angular momentum is exclusively confined to quantized vortices and cannot reside 
in the bulk of the condensates of wave CDM. 
Therefore, the crucial point is: when $T \ll T_\mathrm{c}$, 
a substantial condensate fraction (or more precisely, the superfluid component) must be present in any wave CDM, 
and the gravitational collapse of this curl-free component 
is the desired/dream ``coherent collapse''\cite{Silk-Rees1998}.

Mentioned in passing, 
in those baryonic gas-based scenarios without merger driving, 
a further challenge is against fragmentation, which requires both metal-free gas 
(free of metal contamination from star formation) 
and ultraviolet radiation to prevent molecular hydrogen cooling (thus, not far way from star formation;
cf. \cite{JiaoBrandenbergerKamali2025} for a different proposal about the source of ultraviolet photons).

%MNRAS 2023 : 
%Direct collapse of exceptionally heavy black holes in the merger-driven scenario
%Lorenz Zwick,1 Lucio Mayer,1 Lionel Haemmerle 2 and Ralf S. Klessen
%ApJ 2024 : 
%Direct Formation of Massive Black Holes via Dynamical Collapse 
%in Metal-enriched Merging Galaxies at z~10 -- Fully Cosmological Simulations
%Lucio Mayer , Pedro R. Capelo , Lorenz Zwick , and Tiziana Di Matteo

%% Inayoshi_Visbal_Haiman_2020ARA&A__Assembly of the First Massive Black Holes 

%%%
% PHYSICAL REVIEW D 100, 063507 (2019)
% Formation, gravitational clustering, and interactions of nonrelativistic
%   solitons in an expanding universe
% Mustafa A. Amin 1,* and Philip Mocz 2,  
%  1 Physics & Astronomy Department, Rice University, Houston, Texas 77005, USA
%  2 Department of Astrophysical Sciences, Princeton University,

\section*{Solutions to the false obstacles for wave CDM collapsing to BHs}

As mentioned above, with increasing studies recently on wave CDM (or called fuzzy, ultralight bosonic DM),
there is another possibility that 
seed BHs or even SMBHs themselves are created 
from gravitational collapse of wave CDM condensates
during a certain nonlinear evolution stage of cosmic large-scale structure.
This proposal, if confirmed, implies that 
the mysteries of DM and SMBHs are the same.

However, previous attempts proposing wave-born DCBH scenarios 
were generally stuck by two seeming obstacles,
which are actually misconceptions to be analyzed below.
Then we point out that several recent studies have revealed 
the instability mechanism of spinning boson stars 
(i.e., the hypothesized relativistic version of the solitonic cores of wave CDM haloes), 
which result in black holes (see \cite{Sanchis-Gual2019} and the subsequent investigations\cite{Siemonsen-East2021,Siemnonsen-East2023}).\\

\textbf{Misconception 1: the ``maximum stable mass'' for boson stars is the necessary condition for the wave collapse to BHs.}  \\
%1. the $M_\mathrm{max}$ for static (stationary) boson stars: 
% a conceptual (heuristic) roughly sufficient condition for BH formation, not a necessary condition.
%
%%//give a definition of boson stars here. hypothesized objects \\
BECs and hypothesized boson stars both are  
\rev{macroscopic coherent states %%of quantum particles -- bosons.
with a single phase shared by a macroscopic number of bosons,\cite{Ferreira2021}
and belong to the broad category of so-called coherent structures (neither quantum or classical).\cite{Nazarenko2011}
}
Nothing but the overall phase depends on time, 
with a single frequency (hereafter called the coherent frequency);
hence often in those research fields the two kinds of objects  
are roughly called solitons.

Almost all serious analyses (see, e.g., \cite{Chavanis2016}) 
employed the following condition: a solitonic core of wave CDM (namely a ground-state BEC or Newtonian boson star)
must be more massive than the maximum stable mass (the so-called Kaup limit\cite{Kaup1968}), 
as the criterion of collapse to BHs.
For fiducial parameters of ultralight DM (boson mass around $10^{-22}$ eV 
and ultraweak self-interaction\cite{Hui2021,Ferreira2021}),
the maximum stable mass of the solitonic core (namely the Bose-Einstein condensate with momentum close to zero)
is $\approx 8 \times 10^{11}$ solar mass,
which means that only a tiny fraction of DM solitonic cores in the actual Universe 
could collapse into black holes 
%%\revMay{ }
according to the tight scaling relation between the solitonic core and the hosting DM halo 
(e.g., ref.\cite{Schive2014PRL}; but cf. refs.\cite{YavetzLiLam2022,MaySpringel2023} for significant scatter 
or even non-unique correspondence between the two);
see a similar case in ref.\cite{Helfer2017}. %%2024Nov22
Yet, this maximum-mass condition is actually not necessary.

%% new paragraph: 2025Mar
The ``maximum stable mass'' is established in the line of reasoning as follows.
First of all, a stationary-state solution is assumed and defined for boson stars,
with nothing but the coherent-state phase of all bosons 
being time dependent and oscillating 
(the so-called \emph{stationary ansatz}\cite{Kaup1968}; %% recalling Bohr's stationary states of Hydrogen atoms).
recalling a standing wave).
Then, after general-relativistic calculations and perturbative stability analysis, 
it is found that there exists a maximum mass formula 
beyond which no \emph{stable configuration} is possible for boson stars.

%% new paragraph: 2025Mar
A crucial point lies in the very beginning of the above reasoning, 
that is, boson stars are \emph{assumed to be in static equilibrium of configuration} 
(except for %\revMay{}
the on-site oscillation in time of every boson with the common coherent frequency mentioned above),
which means there is no motions at all in the configurational body of boson stars.
%% (i.e., no any velocity terms for the bosonic fluids).
This point is completely different from the realistic cosmic situation of bosonic fluids such as the dark matter,
particularly in the nonlinear stage of cosmic structure formation.
During the formation processes of large-scale structure and haloes, the bosonic fluids of dark matter
are dynamic and turbulent (not in stationary state or equilibrium at all), 
with all kinds of fluid velocities.

In fact, the Kaup limit is essentially a heuristic argumentation with analytical simplicity,
a concept mainly for the purpose of stability analysis of already-in-equilibrium boson stars
(similar to the concept of Jeans mass used in the situations of normal matter in the weak gravity regime).
\revMar{Kaup limit and Jeans mass 
apply only under near-equilibrium conditions where perturbation analysis is justified,	
e.g., the linear stage of cosmic large-scale structure formation.
Thus both concepts are not applicable in the context of the deeply nonlinear stage of structure formation,
not to mention any collapse process.} 
%% For the judgment of BH formation, 
In the context of black hole formation, 
Kaup limit can be regarded as a roughly sufficient condition,
but definitely not a necessary one.
For example, imagine such a case: all the bosons in the solitonic core
have velocities inward in a coherent manner.
%% 2025May27:
\revMay{A comprehensive investigation of such cases would require advanced mathematical frameworks, 
	such as the wave turbulence theory for the Schrodinger--Poisson system (i.e., wave CDM haloes),\cite{SkippLvovNaza2020}
	which is beyond the scope of this article. 
	Certainly, breaking the Kaup limit through specific velocity fields is merely the first step 
	and does not guarantee black hole formation; see the subsequent discussion on instabilities for further details.}
%%J.Skipp_V.L'vov_S.Nazarenko_2020PRA__Wave turbulence in self-gravitating Bose gases and nonlocal nonlinear optics; 
%%SPE,SHE,NLSE
In short, anyway, the very presence of dynamic processes (instead of static balance) 
is the solution to this seeming obstacle.
\\
%%\commentMay{Xinyu: 1. check $M_\mathrm{max}$ 8e11 on p.5; ~~ 
%% 2. to add comment on the relation from $M_\mathrm{core}$  to $M_\mathrm{halo}$.  --2024.5.13}

\textbf{Misconception 2: ``bosenova'' arrests the wave collapse.}  \\
No ``bosenova'' phenomenon happens at all in the wave CDM case.
Just as explained in the cold atom experiments (e.g., \cite{Donley2001}), 
during the process of BEC collapse the explosion happens 
just because there is force between atoms.
The interactions (or called collisions) among two or more atoms 
lead to several mechanisms that produce and transfer kinetic energy to part of the BEC atoms.
A main mechanism is three-body recombination: during the collision of three atoms,
two of them form a ``molecule'' and the third one is kicked out.
Instead, for the part of wave CDM, 
no force (namely the so-called self-interaction) between bosons is required.
Mathematically, the three-body recombination of bosons here is the same process as in 
gravitational collision of three stars 
leading to the formation of a stellar binary. 
But there is a crucial difference:
while gravity plays the important role in collisions among such \emph{compact} celestial objects as stars,
it does not in the case of wave CDM,
because ultralight DM bosons are \emph{de~Broglie waves} on spatial scales $\lesssim 1$ kpc.

All in all, the most fundamental reason concerning ``bosenova'' is this:
both cold atoms in the laboratory experiments and the above-mentioned normal stars
exhibit their particle-ness,
whereas in the galactic-scale situation of wave CDM, bosons behave as waves. 
\newline

\textbf{Key: Non-axisymmetric instability of spinning boson stars, the formation mechanism of DCBHs.}  \\
Over the past five years, there has been a major breakthrough in the field of boson stars
(namely, the stationary scalar fields bound by their own gravity)  
both in the general-relativistic (GR) and the Newtonian regimes.
Numerical simulations found that 
a dilute scalar cloud can evolve to form a spinning boson star (i.e., with non-zero angular momentum),
the presence of rotation of the boson star then 
triggers non-axisymmetric instability (NAI),
and thus the boson star can collapse to form a black hole 
 \rev{(or put safely, 
a compact relativistic object unresolved at the spatial resolution of 
their simulations).}%2025Feb20 added.
\cite{Sanchis-Gual2019, Siemonsen-East2021,Siemnonsen-East2023} 
The NAI in the GR regime progresses rapidly. 
In the weak gravity limit, Newtonian boson stars (precisely speaking, the solitons with their vortices) 
are unstable marginally\cite{Siemonsen-East2021}, 
but in principle this progression can be accelerated by dynamic processes 
such as interactions and even mergers among boson stars\cite{Siemnonsen-East2023}, 
and/or interactions with their surroundings roughly on dynamical time scales 
(cf. the above solution to the first obstacle; 
\revMay{e.g., a simple case discussed in ref.\cite{Chavanis2016}: 
	the merger of two or more wave DM cores resulting in a soliton surpassing the Kaup mass limit}).

As we know, Newtonian boson stars essentially represent the solitonic cores of wave CDM haloes,
and relativistic boson stars are the same as the DM cores that evolve into the GR regime.
\revMay{Within the hierarchical structure formation scenario of generic CDM haloes, 
a halo acquires angular momentum in the course of its assembly. %%xbdong 2025May24 
Therefore, the above findings appear to perfectly provide the collapse mechanism for DCBHs,
and can be a viable solution to the above Puzzle~2.}
%%%
%% 2025.5.24, 5.26 xbdong adds:  
\revMay{Regarding specific questions -- such as which types of wave CDM haloes may ultimately evolve into DCBHs, 
and the fraction of such haloes at a given redshift -- 
first of all there is indeed a critical gap yet to investigate thoroughly:  
how often and how soon for the aforementioned dynamic processes
to drive haloes from the Newtonian to strong-field gravitational regime. 
In practice, this is a difficult task in terms of the computational limitations of numerical simulations: 
%%for a given spatial resolution, the maximum resolvable velocity is constrained 
%%and may instead not be able to match the required velocities 
%%in the evolution of those processes (e.g., collisions and mergers) 
%%with increasing accelerations. 
%%This mismatch would lead to wrong simulation results. 
%%In other words, %%% -- 2025May27. 
with increasing accelerations the velocities and de~Broglie wavelengths of the bosons increase, 
demanding ever-increasing spatial resolutions. %% in numerical simulations.
%%% --- 2025May27.
Once the systems reach the stage where the monotonically deceasing NAI timescale gets sufficiently short, 
answers can be practically obtained from numerical simulation results such as the ``remnant map'' 
formulated in ref.\cite{Siemnonsen-East2023}.
}
%%\commentMay{Xinyu: to add some words describing the dynamics picture for DCBHs: 
%% e.g., what kinds of waveDM haloes can evolve to DCBHs, how many (fraction of) waveDM haloes 
%% can/have(will) become DCBHs? --2024.5.13}

%F.Di Giovanni_Sanchis-Gual, Nicolas_etal_2020PRD__Dynamical bar-mode instability in spinning bosonic stars
%More precisely, we found that scalar boson stars are affected by a nonaxisymmetric instability 
%which triggers the loss of angular momentum and 
% the reshaping of the energy density profile from a toroidal shape into a spheroidal one. 
%
%Siemonsen_William.East_2021PRD__Stability of rotating scalar boson stars with nonlinear interactions 
%

%% \paragraph{Difference: waveDM-born BHs vs. primordial BHs}
%% Just a sentence,  to mention in passing (in the beginning). 
%%\bigskip
%%%% -- done.

\section*{Inside-to-Outside star formation: in and around the accretion flows toward MBHs}

In the above DCBH scenario, MBHs (the heavy seeds) are born in or near the bottoms of 
some gravitational potential wells of wave CDM.

Then, straightforwardly, 
the next scenario of a special starburst mode 
is just as described in \cite{Silk-Rees1998} and \cite{Collin-Zahn1999}, 
including the following ingredients:
cold gas being accumulated in the local gravitational wells by the DM and the MBHs, 
accretion flows toward the MBHs, 
star formation in the self-gravitating part of the accretion flow 
(the so-called accretion-modified SF, on the scales from the broad-line region to the torus),
and star formation triggered by shocks 
due to both the above accretion-modified SF\cite{Collin-Zahn1999}   
and the AGN outflows\cite{Silk-Rees1998,Silk2024}.
%%% below added after Hai.FU's comments:
This scenario has three special characteristics:
accretion-modified star formation\cite{Collin-Zahn1999,Ali-Dib-Doug2023,JMWang2023}, 
suppression of negative stellar feedback due to high gravitational acceleration\cite{Silk2024,Boylan2024},
and positive feedback of shocks\cite{Silk-Rees1998,Collin-Zahn1999,Silk2024};
see the mentioned references for the details.
%%%
Here we would like to highlight a recent progress\cite{Ali-Dib-Doug2023}
that further investigated in detail the process that stars can be born and die in the accretion flows,
which excellently accounts for 
the observed super-solar $\alpha$ element and Fe abundances in the AGN BLRs\cite{Huang-Doug-Shields2023,JMWang2023}.

%% new para:
Note that the above is an inside-to-outside star formation scenario based on physical mechanisms,
which can account for the descriptive ``inside-out growth'' of galaxies 
(see, e.g., \cite{vanDokkum2010}), 
but not the same, because the routes of 
\rev{ \emph{``growth''} there\cite{vanDokkum2010} (vs. ``star formation'' here) } 
compass galaxy mergers, even dry mergers.  
Thus here we deliberately use the a little bit verbose term ``inside-to-outside'' 
instead of the popular ``inside-out'' in the literature,
also for both the sake of clarity and the exact meaning of directional progression.

%% para:
The most important value of this accretion-flow based scenario of ``inside-to-outside'' star formation 
lies in the fact that 
it is the only theory so far capable of explaining the spatial difference in metallicity 
between the AGN BLR and NLR (NLR being the ISM of the host galaxies)\cite{Huang-Doug-Shields2023};
i.e., the most reasonable answer to the above Puzzle~3.
In fact, this is the very motivation of all the aforementioned studies on star formation 
within self-gravitating accretion disks or in the vicinity of accretion disks, 
dated from \cite{Silk-Rees1998, Collin-Zahn1999, Hamann-Ferland1999} to \cite{JMWang2023, Huang-Doug-Shields2023};
\revMay{it is straightforward to expect that this star formation channel should operate also 
in the high-redshift AGNs that JWST discovered.}

In a certain sense, the stellar clumps resulted from the above AGN induced SF 
\revMar{can be regarded as, in a significant fraction at least,}  
\textbf{the first building blocks of galaxies}.

%\revMar{
%%% Mar 12, adding a new section, after the comments from the Mar3 video meeting:
%%-   George: luminosity function; Peter: power spectrum, of the DM.
%% Mar12: \section*{Wave CDM's power spectrum vs. Observations}
\section*{Prediction for halo mass functions of wave CDM} %2025Mar31
%}

\ReasoningChainMar{
\noindent
Plan (Chain of Reasoning):\\
//Mar 12, adding a new section, after the comments from the Mar3 video meeting (George: luminosity function; Peter: power spectrum, of the DM).\\
1. the most important, first of all (WHY we can do BETTER than other waveDM papers?)\\
PHYsics: all the things such as granules in a waveDM halo, constructive fringes (even within the cosmic filaments)
ARE the SAME as the central condensate physically (the BEC theory --- Hartree--Fock--Bogoliubov )!! \\
\{only a minor difference, but not a qualitative one: the extent of containing elementary excitations (phonons, vortexes, collective modes),  
more or less.\}

2. In this article, we handle the DM part only, NOT cover the baryon--DM
(e.g., don't talking about things blablabla; becoz TOO Technical!).\\
That is, talking only about the enhancement mechanisms at small scales  due to wave CDM per se!
(see point 1: various types of condensate structures in a waveDM halo).

LuminFuncts: concerning galaxies. if relate LFs to dark matter's halo mass function, then we need messy things: baryon--DM, galaxy formation. Too technical at this point.

PowerSpecs: the observations concerning tracers only, such as Ly-alpha absorbing gas.
Tracers (tracing DM) are simpler.

3. Be concrete: the ``condensates" (and their gravitational peripherals): the ABC knowledge of BEC , de Broglie waves (physics department): \\
the sizes, living time-scales (their lower limits), etc., of granules and bright fringes in a halo: \\
coherent time: wavelength\_deB / v; == de Broglie time-scale (freq.): h/E = h / (0.5 m * V$_\mathrm{viral}^2$), phonon's energy unit; \\
de Broglie wavelength, namely a granule's size: h/p = h (m * V$_\mathrm{virial}$ ), i.e. the essence: 
the same order as the soliton core!

4. Our final strike -- to get the real thing: \\
2025Mar12: estimate a power spectrum at, say, z = 14(?). \\
 $[$ 2025Mar25 Note: because of the limited data from the simulations, change to plot halo mass functions of waveDM. $]$   \\
Way out: best to be on the shoulders of the simulation guys. \\
 Todo list: to survey the best simulations of waveDM ( and / or warm DM also? ) , best so far.

4.1:
Schive+2016: particle simulation for waveDM LSS. \\

4.2: 
Mocz+2020 (becDM paper 2);

4.3: May and Springel 2023: SPE simulation (spectral method).

Mocz+2020: w/ baryons, pointing out: first galaxies originate in the condensates of cosmic filaments.

Power spectra in May and Springel for all redshifts;
halo mass functions: only for z=3 in May and Springel 2023; a formula for all redshifts in Schive+2016. \\
{\color{red}
---$>$  Their limitation: \\
 A soul-problem: ``What the hell is a halo?", they only count the round, well virialized objects.  \textbf{But the point is: what are the DM objects for galaxy formation (baryon accumulation and star formation)?} }

Then, choice: Power spec vs. halo mass function:\\
The philosophy of this article: 
To separate/isolate/minimize the complex galaxy-formation effects from our consideration (as much as possible)! \\
\{//the philosophy of QFT, namely e.g. Schwinger and effective field theory:
cut off / seperate the uncertainties what we can't know/test at this point 
 from what we can consider based on JWST's observations. \} \\
  $[$ 2025Mar25 Note: because of the limited data from the simulations, change to plot halo mass functions of waveDM. $]$ 

//The above plan is the chain of reasoning during 2025Mar5 -- Mar12. \\

\noindent //2025Mar25 continues (after the long pause owing to the March proposal season):

The simulations: \\
4.4:
Nori, Murgia, Irsic,Baldi and Viel 2019,  Lyman-alpha forest and non-linear structure characterization in Fuzzy Dark Matter cosmologies; same as their 2017PRL paper; 
Madelung SPH code for waveDM.
(Strange thing in those two papers: surplus power than the CDM PS in the k = 0.8--1.2 /Mpc range. This should be a technical problem. )

4.5:
the so-called ``classical FDM'' simulations, T.Dome series.
Dome+2023 a\&b: has PSs at different redshifts, no halo MFs (albeit nodes, filaments, walls in Dome+2023b). \\
Yet the problem is: this method didn't handle quantum pressure term in the evolution dynamics, let alone wave interference.

4.6: other simulations: no useful data for our purpose. (zoom-in simulations are for a single halo, 
while the later-on large-scale simulations (after May and Springel 2023) have to use perturbation techniques, etc. 
(i.e., can't handle THE Deeply NonLinear stage).

//\textit{ ---- end of Part 4 of the reasoning (the literature summary of the simulations data / figures).}\\

5. Decision:\\
In light of the data and figures in the above-listed simulations, only a few papers 
have useful information about classical haloes and other types of condensate objects.
The most useful information: May and Springel 2023 have assess the numbers and mass functions of classical haloes (round, virialized objects), 
and those of granules and dense interference patterns in cosmic filaments!\\
So, it's straightforward for us to plot halo mass functions, instead of plotting power spectrum. \\
--- 2025Mar25, end of the chain of reasoning.

 } %% 2025March12. %% 2025Mar25, end of the reasoning.

%\bigskip
%%\revMar{ 
%\noindent
%//\textit{(Mar12 -- formal text will begin here, blablabla:)}
 %% - restarts on 2025Apr14 (based on the Apr2 circulated version):

%% 2025Apr18, choose the version with GSMF, and thus comment out the following figure:
%\begin{figure*} %% Figure 2:
%	\centerline{\includegraphics[scale=1,angle=0]{Figure2_HMFs_wMaySpringelCurve_idl.pdf}}
%	\caption{\textbf{Illustration of the halo mass functions (MFs) of wave CDM.}
%	The MF of nonclassical haloes (blue dashed line) of wave CDM is obtained by fitting the model (Eqns.~1 and 2) to 
%	the Schrodinger--Poisson wave simulation result with the particle CDM initial condition of May \& Springel (2023)\cite{MaySpringel2023}
%	(light blue, dotted line), for wave CDM at redshift $z = 3$ with boson mass $mc^2 = 7 \times 10^{-23}$~eV.
%	The MF of classical haloes (red dashed line) is well represented by Equation~2, as found and confirmed by refs.\cite{Schive2016,MaySpringel2023}.
%	The total MF of wave CDM (cyan) is plotted in comparison with the particle CDM counterpart (the Sheth--Tormen form, black line).
%    }
%   \label{fig:HMFs}
%	%%\label{Figure2}
%\end{figure*}

%%%%%%%%%%%%%% 2025Apr2, moved here:
\begin{figure*} %% 2025Apr18 choose this version, with GSMF, to be Figure 2:
	%% in the IDL plotting code:
	%%     ;; GSMF: d n / d logM ; but Halo MF is expressed as d n / d lnM.
	%%     ;; so, GSMF whoud be boosted by GSMF = GSMF * ln10. -- xbdong, 2025.4.14
	%%     MF_NC24 = MF_NC24 * alog(10d)  ; Navarro-Carrera2024 et al. ApJ
	%%     MF_TNG100 = MF_TNG100 * alog(10d)
	%%
	\centerline{\includegraphics[scale=1.8,angle=0]{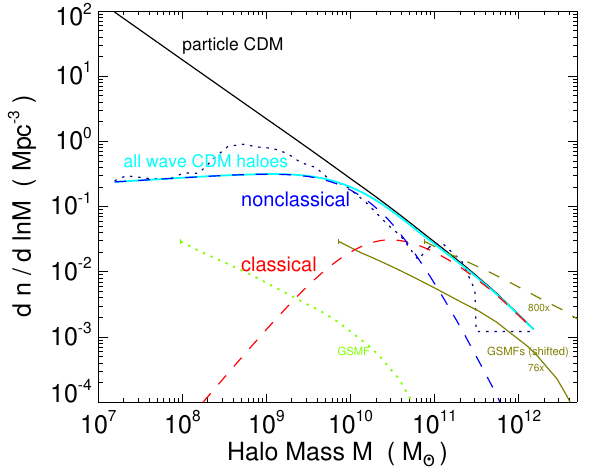}}
	\caption{
	 \textbf{Illustration of the theoretical halo mass functions of wave CDM, confronted with the observational galaxy stellar-mass function.}
    %% The theoretical MFs of haloes at $z=3$ are the same as in Figure~\ref{fig:HMFs}.
	The MF of nonclassical haloes (including subhaloes; blue dashed line) of wave CDM is obtained by fitting the model (Eqns.~1 and 2) to 
	the Schrodinger--Poisson wave simulation result with the particle CDM initial condition of May \& Springel (2023)\cite{MaySpringel2023}
	(light blue, dotted line), for wave CDM at redshift $z = 3$ with boson mass $mc^2 = 7 \times 10^{-23}$~eV.
	The MF of classical haloes (red dashed line) is well represented by Equation~2, as found and confirmed by refs.\cite{Schive2016,MaySpringel2023}.
	The total MF of wave CDM (cyan) is plotted in comparison with the particle CDM counterpart (the Sheth--Tormen form, black line).
    %%%
    The green dotted line shows the galaxy stellar mass function (GSMF) at $3.5 < z < 4.5$ from Navarro-Carrera \etal\ (2024).\cite{Navarro-Carrera2024_GSMFs}
    The dark yellow lines show the corresponding Halo MFs inferred from the GSMF by assuming $M_\mathrm{halo} = 76 M_\mathrm{\star}$ (solid) and 
    $M_\mathrm{halo} = 800 M_\mathrm{\star}$ (dashed), respectively.
    Note that the x-axis of the GSMF (green curve) corresponds to stellar mass ($M_\mathrm{\star}$) and 
    the lower bound probed so far is around $M_\mathrm{\star} = 6 \times 10^7$\msun. 
    }
   \label{fig:HMFs_vs_GSMF}
	%%\label{Figure2}
\end{figure*}

%Paragraph 1: about the physics (waveDM core condensates and quasicondensates/granules/constructive ridges/... are the same. 
%classical haloes and nonclassical haloes); should be in a friendly style for astronomers. 2025Mar31: to be added later.\\
%... ...

\revApr{
A striking phenomenon common in classical fluids, plasmas and quantum fluids 
is the spontaneous emergence of coherent structures (no matter whether classical or quantum ones)
out of incoherent nonlinear wave fields; as a particular type, condensates result from 
the so-called inverse cascade of particles toward long-wavelength end ($k \rightarrow 0$).\cite{Nazarenko2011}
In the case of wave CDM haloes, 
because the temperature of the bosons is far below the critical ($T/T_\mathrm{c} \rightarrow 0$),  
the entire system with excited higher-energy states 
can be well described in terms of the macroscopic wavefunction (namely the order parameter) and its fluctuations.\cite{GriffinBook1993} 
This means that granules, filamentary cores and ridges in the outskirts of wave CDM haloes and in large-scale structures (e.g., as subhaloes)%
\cite{Hui2021,Mocz2020,MaySpringel2023} are equivalently condensates,
albeit with more fluctuations in their macroscopic wavefunction than the exact ground state 
(presumably the central solitonic cores of the haloes and subhaloes).
In other words, they are all the same physically, 
except for differences in the amount of collective excitations such as vortices and phonons they contain.\cite{GriffinBook1993} 
%%  Bogoliubov-Beliaev-Popov theory; 2025Apr14
This is analogous to ripples on a lake surface: 
ripples and ripplets --- regardless of their specific morphology or size ---  
are fundamentally of the same physical nature as standing waves (sometimes even called ``solitons'' in the literature).
} %%\revApr{}

Recently, May \& Springel (2023)\cite{MaySpringel2023} have found a large proliferation of low-mass objects of high densities 
--- in additional to classical haloes (approximately spherical and virialized) --- in their wave CDM simulations,
which are not numerical artifacts, but have two physical origins (see also ref.\cite{Mocz2020}): 
``localized density fluctuations arising from the wave nature'' (i.e., constructive patterns of wave interference) 
and  ``high-density filamentray structures'' (including the ``bright interference ridges'' inside the filaments).
According to the physical rationale outlined above (at $T \ll T_\mathrm{c}$),
those two types of dense objects they found\cite{MaySpringel2023} 
are essentially condensates (albeit being excited to some degree) with their gravitational peripherals,
just like the ground-state condensate in the center of classical haloes.
In light of their high densities\cite{MaySpringel2023}---i.e., deep gravitational potential wells 
to attract and accumulate baryonic gas for star formation---we call those objects ``nonclassical haloes'' hereinafter
(mainly being subhaloes).
All varieties of wave DM condensates share the same underlying physics; for instance, 
their characteristic sizes and coherence times are on the order of magnitude universally set by
the de~Broglie relations.

Furthermore, ref.\cite{MaySpringel2023} had an important observation that  
the Schrodinger--Poisson simulations with the particle CDM initial conditions (ICs) 
``can be viewed as an upper limit for the HMF expected for fuzzy DM ICs 
(with the proliferation of low-mass objects)'' (see their Figures~3 and 9).
This inspires us to take the wave simulation with the particle IC as an order-of-magnitude estimate of 
all the ``haloes and subhaloes'', including the nonclassical ones 
that play the same role as classical ones for galaxy formation as explained in the above.
In this way, based on their simulations for HMFs at $z =3$ (see their Figure~9), 
we estimate the mass functions of the nonclassical haloes of wave CDM as follows
(hereinafter, by ``halo mass function'' the contribution from subhaloes is also included).

%%2025March27:\\
Considering that the dynamics of wave CDM (namely the Schrodinger--Poisson equation) admits a scaling fashion,\cite{Hui2021}
%%% recalling the situation in particle CDM:
%% both the gravitational many-body dynamics and halo's build-up history admit a scaling fashion somehow 
%% (within specific scales and to some extents) 
we posit that the MF of the nonclassical haloes relates to the classical counterpart 
with a power-function term, as follows:
\begin{equation}
	\left.\frac{dn}{dM}\right\rvert_{\mathrm{noncls}}(M,z) =
	\left.\frac{dn}{dM}\right\rvert_{\mathrm{cls}}(M,z) \,
	\left[ C \left( \frac{M}{M_0} \right)^{\kappa ~} \right] ~,
	\label{eq:nonclassicalHMF}
\end{equation}
and the MF of all the wave DM haloes is the sum of the two.
%%%%
As proposed by ref.\cite{Schive2016} and confirmed by ref.\cite{MaySpringel2023},
the MF of classical wave DM haloes (the so-called ``genuine'' haloes in those papers) 
can be well fitted by the following formula:
\begin{equation}
\left.\frac{dn}{dM}\right\rvert_{\mathrm{cls}}(M,z) =
\left.\frac{dn}{dM}\right\rvert_{\mathrm{CDM}}(M,z) \, 
\left[ 1 + \left( \frac{M}{M_0} \right)^{-1.1 \,} \right]^{-2.2} ~.
\label{eq:Shive16-HMF}
\end{equation}
%%%%%
In the above two formulae, 
$M_0$ is the characteristic mass below which the MF of classical wave DM haloes 
starts to drop noticeably, defined by \cite{Schive2016} to be 
$M_0 = 1.6\times 10^{10}\,{m_{22}}^{-4/3} \msun$, 
where $m_{22}$ is the boson mass in units of $10^{-22}$~eV.
By applying this model (Eqns.~\ref{eq:nonclassicalHMF} and \ref{eq:Shive16-HMF}) 
to the Schrodinger--Poisson simulation result with particle CDM IC of ref.\cite{MaySpringel2023} 
(as displayed in their Figure~9),
we find that the proposed power-function term (in Eqn.~\ref{eq:nonclassicalHMF}) works fairly well,
yielding best-fit $C = 3.10$ and $\kappa =-1.43$.
The model MFs estimated for the nonclassical and entire haloes of wave CDM (including subhaloes; see above), 
together with the classical one (the formula of ref.\cite{Schive2016}) and the particle CDM MF (Sheth--Tormen form),
are illustrated in Figure~2.
%% (see the Appendix for the detail).

%%% 2025Apr17, move GSMF to here:
\revApr{
In order to observationally test the above theoretical MFs of classical and nonclassical haloes of wave CDM, 
we derive two HMFs from the observational GSMF at  $3.5 < z < 4.5$ by ref.\cite{Navarro-Carrera2024_GSMFs} (the green dotted line).
One HMF (the dark yellow, solid line) is obtained in terms of the cosmic averaged stellar-to-halo mass ratio,
namely $M_\mathrm{halo} / M_\mathrm{\star} = 0.2589 / (0.07 * 0.0486) = 76$; this is the absolute lower-limit HMF constrained by their observed data.
The other one (the dark yellow, dashed line) is obtained by simply assuming $M_\mathrm{halo} / M_\mathrm{\star} = 800$, in order to roughly match 
the theoretical HMFs (no matter neither the particle CDM one or that of all wave CDM haloes),
and could be regarded as a reasonable upper-limit HMF constrained by the observations.
Thus, any correct theoretical HMFs should be consistent with the observationally allowed region between the two dark yellow lines.

Note that the low bound of the GSMF probed observationally by ref.\cite{Navarro-Carrera2024_GSMFs} 
is at $M_\star \approx 6 \times 10^7$\msun. 
Thus, an immediate inference is that if the DM is indeed ultralight, then the MF based on classical haloes and subhaloes 
appears not enough to account for the observed GSMF 
(certainly, careful analysis of measurement errors remains to be conducted to definitively substantiate this conclusion).
Looking ahead, we anticipate significant breakthroughs from upcoming JWST campaigns. 
Should the GSMF at $z \approx 3-4$ be probed by one magnitude fainter than current limits, 
this would enable the firm lower-limit to HMF down to $M_\mathrm{h} = 2 \times 10^9$\msun\ (cf. the dark yellow, solid line).
This enhanced sensitivity will resolve the critical divergence in the HMF shape (cf. the cyan vs. black solid lines)
between particle-like and wave-like CDM scenarios. 
} %%2025Apr - GSMF moved here %%\revApr{}

%% --------  2025Mar12: move the following para. to this new section:
%%para: Answer to Puzzle 1:
%[MEMO -- TBD: The anser to puzzle 1: the enhanced ``localized gravity''. (cf. Boylan2024) \\
%2407.10900 Boylan-Kolchin mnras Accelerated by Dark Matter - 
%a High-redshift Pathway to Efficient Galaxy-scale Star Formation :\\
%providing the requisite accelerations on the scale of entire galaxies in the early cosmos. 
%The key insight is that characteristic accelerations in dark matter haloes ... \\
%]

\bigskip
To conclude, regarding the solution to Puzzle~1, 
there are mechanisms almost equally important to enhance ``localized gravity'' at least on two levels.
On the level of DM \textit{per se}, 
{the ground-state condensates and condensate-like objects (excited to some degree) 
of wave CDM --- in various types of morphology such as 
the central solitonic cores, cylindrical cores of the cosmic filaments, 
and the so-called granules (namely constructive interference patterns everywhere) ---} 
act as the cores of deep gravitational wells, 
which in turn become sites to form DCBHs and stars.
As illustrated in Figure~2, 
contrary to prior expectations for wave CDM,
these galaxy-forming nonclassical haloes and subhaloes
exhibit no suppression in the low-mass regime of the mass function, 
although not as abundant as the particle CDM counterparts 
for halo masses below the quantum Jeans mass (i.e., roughly the aforementioned $M_0$).
On the second level, as highlighted in the above sections,  
the MBHs born out of wave CDM 
provide enhanced gravity and then attract ``enhanced baryons'' on sub-galactic scales.
More subtle mechanisms of enhancement, e.g., on the level of interaction between baryons and wave CDM structures,
may be still possible; this is beyond the scope of this article. 
Anyway, in compact\cite{Silk2024} regions of high gravitational accelerations, 
stellar feedback is insufficient to overcome the gravitational potential well,
leading to a special mode of star formation with high efficiency.\cite{Boylan2024}
Furthermore, refs.\cite{Collin-Zahn1999,Silk-Rees1998,Silk2024,Ali-Dib-Doug2023,JMWang2023} present 
the detailed mechanisms and processes of this star formation scenario.
%%% below, 2025Feb20 added:
\rev{
	We noticed that a recent paper\cite{ChiuSchive2025} proposed that 
	the solitonic cores of wave CDM  in high-redshift galaxies 
	can boot the accretion rate of baryonic gas toward the central SMBHs,
	essentially because of the same role of providing additional gravity as described here for star formation.
	Note that ref.\cite{ChiuSchive2025} did not address the formation of seed BHs,
	but just assumed a pre-existing heavy seed of $10^5$\msun.
}%%ChiuSchive2025

%% } %\revMar{}
 
%\cleardoublepage

\bigskip
%- - - - - \\
%In summary:
\section*{Summary and next steps}
In a nutshell, 
by combining the two old but still radical ideas/sparks 
%% (DCBHs from wave CDM haloes $+$ star formation near the DCBHs)
(DCBHs from ``coherent collapse''\cite{Silk-Rees1998} $+$ star formation near the DCBHs\cite{Collin-Zahn1999,Silk-Rees1998})
and a recent progress in the GR instability of spinning boson stars\cite{Sanchis-Gual2019},
as far as the observational phenomena are concerned,
all the JWST puzzles so far can be explained.
``Heavy eggs (seeds of MBHs)'' precede and
accelerate the growth of ``massive chickens (galaxies)'' in the cosmic dawn.
%%%% 
Yet, for the part of physics, this is just the start of a long journey:
to explore all the concrete mechanisms in the processes 
from the wave collapse of self-gravitating bosonic fluids 
(generally, the single- and multi-component Schrodinger--Poisson systems)
into the general-relativistic (GR) regime,
particularly to understand quantitatively the GR instabilities leading to the BH formation
(e.g., the aforementioned NAI for bosonic solitons with non-zero angular momentum).
%%%%
Ultimately, this may lead us to the following possibility: 
the three puzzles --- the nature of dark matter and dark energy 
(because the dark two are probably connected\cite{Niemeyer2020,Ferreira2021,Hui2021}) %2024Nov22   
and the origin of supermassive black holes --- are the same thing.

On the side of astronomical observations,
the next ten years are particularly promising,
with several large facilities coming online and able to test 
the predictions of both the present proposal 
and wave CDM (the basement of our theory) in general.
As stated above, compared with particle CDM, 
the effect of wave CDM on galactic and sub-galactic scales is unique and two-fold: 
\emph{both enhancing the overall gravity and blurring the detailed structures of matter 
	on those so-called small scales.}
%To test wave CDM predictions, several observational strategies are particularly promising. 
% (1)
High-resolution spectroscopic observations from JWST NIRSpec IFU allow us to examine 
the internal structures and dynamics of high-redshift galaxies, 
providing insights into the ``inside-to-outside'' star formation activities
initiated by wave-born DCBHs. 
%By studying the influence of accretion flows on star formation and quenching within these galaxies, 
%we can assess whether the DCBH scenario aligns with observed galaxy evolution. 
%% (2)
Complementing the above detailed studies within galaxies,
 large-scale galaxy surveys from JWST, EUCLID, Roman (including its synergy with Subaru PFS and Ultimate), 
 and Rubin 
will assemble large samples of high-redshift galaxies, 
enabling statistical analyses of galaxy and SMBH evolution over cosmic time. 
These datasets can probe whether the rapid early growth of massive black holes and galaxies matches wave CDM's predictions. 
%%% (3)
Moreover, constraints on the small-scale matter power spectrum, 
particularly down to the smallest scales ($\lesssim$100 kpc) 
that can be probed by Lyman-$\alpha$ forest observations in DESI and DESI-II,
will provide crucial tests of wave CDM 
by measuring density and velocity fields of dark matter. 
%% Such measurements will constrain the energy scale of wave dark matter. %YD 2024Nov18
%the mass and effective temperature of wave dark matter. $YD 2024Nov10

%%% end of the Main Text. %%%

%%Reference
\bibliography{waveBH}% common bib file
% OR:
%\begin{thebibliography}{}

%\end{thebibliography}

\begin{addendum} 
	
	\item[Acknowledgements]
	We particularly thank Shude~Mao for his warm encouragement and support. 
	We are also grateful to Tim Eifler, Hai Fu, Luis Ho, Zhiyu Zhang, Huiyuan Wang and Zhen Pan for their constructive comments on the manuscript.
	%%%%%%fundings:	
	XD acknowledges support from the National Natural Science Foundation of China (NSFC \mbox{No.12373013}). 
YZ, MJR, and GHR acknowledge support from the NIRCam Science Team contract to the University of Arizona, \mbox{NAS5-02015}.	
	
	\item[Author Contributions]
	All authors reviewed and provided critical feedback on the manuscript throughout its development. 
	\mbox{XD} and \mbox{YZ} contributed equally as co-first authors. 
	\mbox{XD} conceived the project, led the theoretical framework development, and drafted the initial manuscript.  
	\mbox{YZ} led the observational content and overall integration with JWST data, 
	and co-wrote observational sections with \mbox{XD}. 
	\mbox{MJR} and \mbox{GHR} contributed to the content on 
	JWST observations, implications and empirical tests of theoretical predictions.
	\mbox{XL} contributed to the content on wave dark matter. 
	\mbox{PB} contributed to the content on cosmological context and empirical tests of theoretical predictions.
	\mbox{HM} and \mbox{RM} contributed to the text revisions and made Figure~1. 
	\mbox{ZM} contributed to the text revisions. 
	\mbox{ZS} contributed to the content on condensed matter physics and validated the theoretical formulation.

	\item[Competing interests statement]
	The authors declare no competing interests.
	
	\item[Correspondence and requests for materials] should be addressed to Xiaobo Dong (\mbox{xbdong@ynao.ac.cn}).

\end{addendum}

\clearpage
%\section*{Background: Why we write this Opinion/Comment essay?}
%\appendix{Background: Why we write this Opinion/Comment essay?}

%\appendix
%% May30: \noindent \textbf{APPENDIX 1}\\
\noindent \textbf{Supplementary Note}\\
\vspace*{-0.3 in}
%%\section*{Background: Why we write this Perspective article?}
\section*{Background: The Motivations for This Perspective}
%% Opinions? Commentary essay ??

In this appendix, we provide an expository account of the motivations behind the writing of this article.

%Physical Sciences \newline
%including theoretical physics (even quantum fields on curved spacetime), 
%condensed matter physics (cold atoms and quantum fluids), optics, 
%fluid mechanics and plasma, 
%observational Astronomy, theoretical astrophysics (particularly galactic dynamics), cosmology ... \\
%% - Almost all the branches of the physics and astronomy disciplines.

%% first of all: the inter-disciplinary nature:
\textit{Motivation I.} The study of dark matter, the origin of supermassive black holes (SMBHs), 
and the formation of first-generation galaxies 
is a highly interdisciplinary field. 
%%% -It integrates multiple branches of physics and astronomy --.
It requires a comprehensive synthesis of knowledge from 
diverse branches of Physics and Astronomy disciplines,
and thus demands extensive cross-disciplinary collaboration.
For this very reason, the narrative style of this text is designed to be 
intelligible to a broad readership in the Physical Sciences.
%%%
This article seeks to catalyze and foster such intellectual integration,
at least helping researchers get past the jargon that often separates those branches.
%% -with a minimum goal to get the researchers past the jargon of those diverse domains.

Take the solitonic cores of wave CDM haloes as an example.
First of all, we need to understand the associated context in Astronomy and Cosmology 
(in particular, including the theoretical knowledge of both cosmic structure formation and galactic dynamics), 
and be familiar with observational phenomena.
(Here we would like to stress that, as a rule of thumb, 
the concrete knowledge of phenomena 
may appear messy and not intellectually challenging, 
but are crucial if one want to address astrophysical problems.)
%%%
In Physics, there is their closest equivalent, called Bose-Einstein condensates, 
in the field of cold atoms and quantum fluids. 
%%% 2025May1 adds --- HFB theory:
\revApr{
Specifically, it is important to recognize that 
at temperatures far below the condensation critical temperature ($T \ll T_\mathrm{c}$,
which is just the physical situation of wave DM haloes),
a system of bosons (no matter neither excited with higher-energy states 
or not) --- including those with gravitational interactions (namely the Schrodinger--Poisson system) --- 
can be effectively described by the macroscopic wavefunction of a condensate and its fluctuations.
That is, only ground-state condensates and collective excitations 
(namely quasiparticles such as phonons, vortices and various somehow localized modes) need be considered
within the well-established Hartree--Fock--Bogoliubov formalism 
(see, e.g., book\cite{GriffinBook1993}; the ``microscopic'' and mathematically powerful framework of the Bogoliubov theory 
was initially aimed to explain Landau's phenomenological theory of superfluidity,
and has subsequently found applications in calculating diverse phenomena, 
including the structure and reactions of atomic nuclei).
} %\revApr{} 2025May1

%% set to be a new para: 2025May1:
Then, from a broader perspective in Condensed Matter Physics and Optics, 
solitonic objects and condensates are manifestations of macroscopic coherent states 
of quantum particles
(or even related to the so-called macroscopic quantum coherence), 
and thus the entire toolkit of quantum kinetic theory developed in those fields 
can be effectively applied to wave CDM.
%%% - wave kinetics; coherent structure --:
From an even broader perspective, solitons and condensates have analogs in Plasma Physics and Fluid Mechanics, 
where they are typically regarded as a particular type of coherent structures 
(although there may be subtle differences in physics, e.g., the origin of integrability vs. ergodicity). 
In these fields, there exists a related systematic theory known as (weak) wave kinetics.
%%%% -- GR: QFT:
On the other hand, when the general-relativistic version of the solitonic cores is considered for the formation of DCBHs, 
knowledge from the field of relativistic stars becomes relevant 
(e.g., the aforementioned GR instabilities in rotating stars). 
Moreover, there is an exact counterpart, called boson stars, 
which are stationary-state scalar fields coupled with their own strong gravity. 
Then, alas, we have to be familiar with a certain knowledge of field theory, even quantum fields, in curved spacetime.
%%%% -----.
After penetrating the jargon jungle and 
grasping the same essence behind those different names 
(as well as being keenly aware of the differences in their concrete physical incarnations),
we could try to integrate the developments of those multiple disciplines 
into a unified theory for the above stated ``cosmic gravitational conundrums.''

%% 2nd: facts vs. speculations, is important! ---
\textit{Motivation II.} Through this review, we hope to help researchers in this field 
distinguish between what are facts and what are speculations. %% -important!!!
Regarding facts, certainly we must accept them exactly as they are 
(e.g., the cosmic total, matter and baryonic density parameters). 
As for speculations, we must remind ourselves: they are not facts, and therefore we should keep an open mind, 
considering the possibility of alternative theoretical explanations 
(e.g., production mechanisms for the ultralight bosonic field). 
  \rev{Certainly we know it is an essential part of the research process
to propose hypotheses (such as working models). 
But the point is that in most cases as the investigation progresses
the initial hypothesis has to be continuously revised --- at times even totally changed;
in any exploration we have to keep an open mind and 
are willing to keep updating or changing the working hypothesis
(wisdom from many creative scientists such as Richard Feynman, Philip Anderson, and Heisuke HIRONAKA). 
}

To illustrate this point, let us examine the topic of
``the origin of SMBHs'' as a case. 
Theoretical researchers in the Astronomy community 
tend to propose scenarios based on directly observable astrophysics, 
essentially focusing on various processes involving baryonic gas\cite{Inayoshi2020}.
On the other hand, researchers on the Fundamental-Physics community love to seek solutions 
by introducing possible ``new ingredients'' in the style of 
the discipline of quantum field theory and elementary particle physics 
(e.g., primordial black holes and some hypothetical particles created in various hypothesized inflation models).
%%%%
Evidently, it would be much better for both sides to communicate with each other 
to determine which parts of their own theories should remain replaceable %%flexible 
and what should be regarded as established by observational facts. 
This dialogue may even lead to the recognition of real essential concepts 
shared in common by the two schools of thought. 
More beneficially, this could help each side identify gaps in their own reasoning, 
and thus potentially open up new opportunities to address the problem.

%//Einstein 1905: Goal:  %% Einstein 1905:
%to explain several puzzling phenomena
%by assuming that 
%$<$the interaction of light with matter consists of the emission or absorption of such light quanta.$>$ \\
% - Einstein constantly reminded his colleagues of the need to introduce 
%\textbf{ radically new concepts } 
%to explain the structure of both matter and radiation. 
%He himself introduced some of these new concepts, notably the light quantum hypothesis, 
%although he remained unable to integrate them into a coherent physical theory.

\elegantrule  %; divider line.

Lastly, on the ``dark matter/field/gravity'' conundrum, we would like to 
hear a theory (or a research program) that might whisper from the walls of evidence:
%\vspace*{-0.08 in}
\vspace*{-0.6\baselineskip}
\begin{center}
	\emph{``Wave CDM stands \mbox{unfalsified} by observations of galaxies, bearing witness to all relevant puzzles thus far.''}
\end{center}

%%% comment out the planned Appendix 2, -- 2025Apr14:
% \clearpage
% \bigskip
%\revMar{ %% 2025Apr2
 %\noindent \textbf{APPENDIX 2}\\
 %\vspace*{-0.3 in}
 %\section*{Brief description of the halo mass functions}
 %2025Apr2: 
 % Formulae used in Fiures~\ref{fig:HMFs} and \ref{fig:HMFs_vs_GSMF}, to be written later.

  %Memo: a technical issue yet to check: Seems that there is a difference by ln10 (shift in y direction)
  % between halo's MFs ( lnM, theorists' hobbit) and galaxy's LFs/MFs (count numbers in log10M bins ). 
%} %%\revMar{ %% 2025Apr2

\begin{figure*}
	\centerline{\includegraphics[scale=0.94,angle=0]{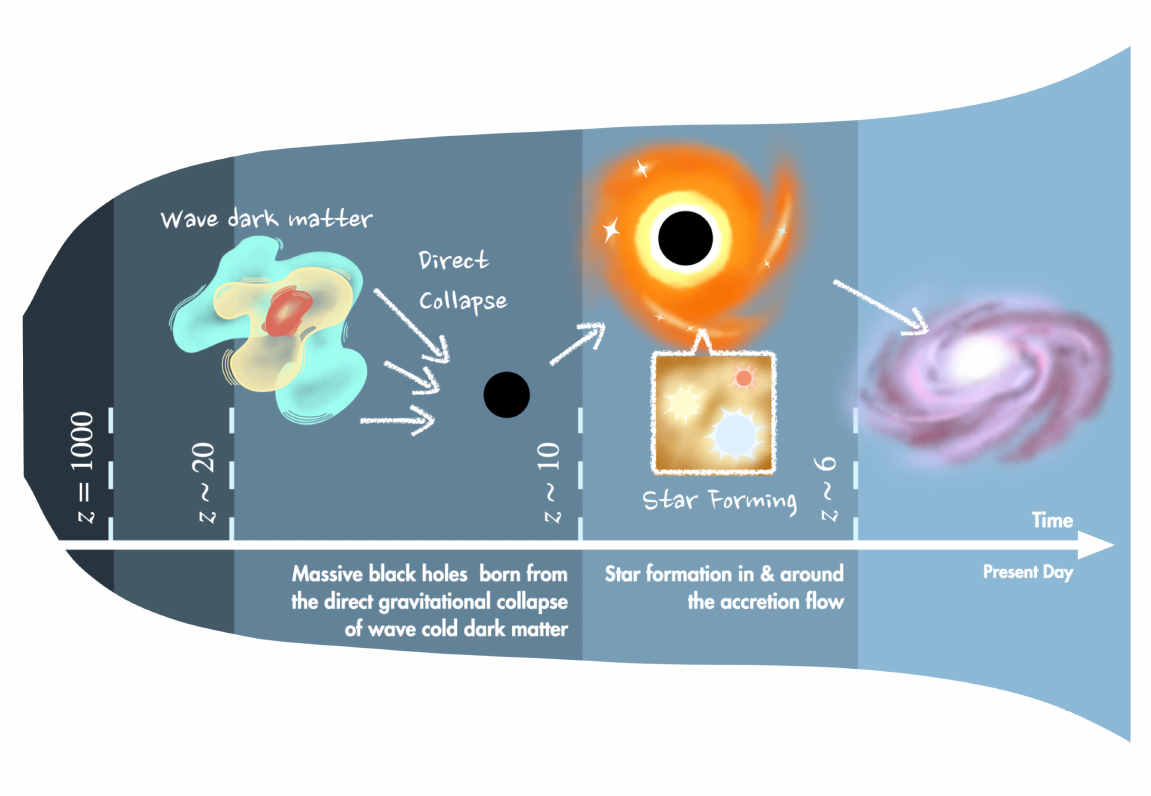}}
		%%{Schematic_2D_Haixia_2024-12-21.pdf}}
	\caption{\textbf{Option 2: Schematic diagram of the theory proposed to explain the JWST puzzles and other ``cosmic gravitational conundrums.''} 
		The following three ingredients are integrated into a unified scenario:
			(1) the formation of massive black holes precedes first-generation galaxies\cite{Silk-Rees1998},
			(2) which are directly collapsed from wave dark matter 
			due to non-axisymmetric instability numerically found in spinning scalar-field boson stars\cite{Sanchis-Gual2019},
	and (3) the inside-to-outside star formation in and around the accretion flows toward massive black holes\cite{Silk-Rees1998,Collin-Zahn1999,Ali-Dib-Doug2023,JMWang2023,Silk2024}.					
        (optional: 2D version, hand-drawn) %(draft ver.1; optional 1: 2D,  hand-drawn)
     }
	\label{Figure1-2D}
\end{figure*}

\begin{figure*}
	\centerline{\includegraphics[scale=0.26,angle=0]{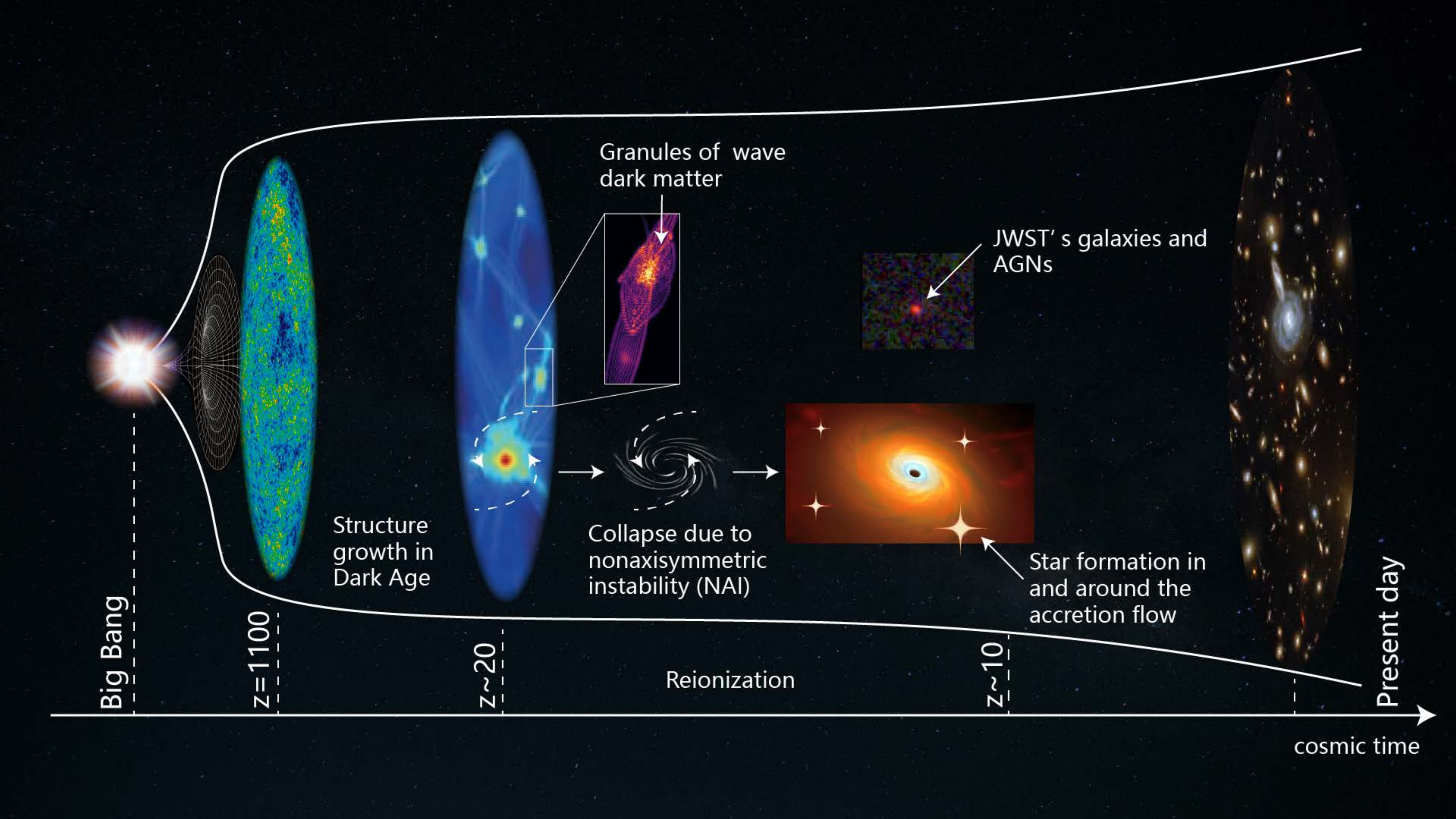}}
	%\centerline{\includegraphics[]{Schematic_3D_Runyu_2024Dec22.eps}}
	\caption{\textbf{Option 3: Schematic diagram of the theory proposed to explain the JWST puzzles and other ``cosmic gravitational conundrums.''} 
	The following three ingredients are integrated into a unified scenario:
	(1) the formation of massive black holes precedes first-generation galaxies\cite{Silk-Rees1998},
	(2) which are directly collapsed from wave dark matter 
	due to non-axisymmetric instability numerically found in spinning scalar-field boson stars\cite{Sanchis-Gual2019},
	and (3) the inside-to-outside star formation in and around the accretion flows toward massive black holes\cite{Silk-Rees1998,Collin-Zahn1999,Ali-Dib-Doug2023,JMWang2023,Silk2024}. 
		Data credits: CMB -- Bennett \etal\ 2013, Astrophys. J. Suppl. 208, 20; 
		waveDM LSS: -- Schive \etal\ 2014, Nat. Phys. 10, 496; 
		waveDM granules -- Mocz \etal\ 2019, Phys. Rev. Lett. 123, 141301;
		HST image: NASA/ESA/Hubble, F.~Pacaud, D.~Coe\,.
        (optional: 3D version) %%(draft ver.1; optional 2: 3D). 
	 }
	\label{Figure1-3D}
\end{figure*}

 \cleardoublepage

\end{document}